\documentclass[12 pt]{article}
\usepackage[utf8]{inputenc}
\usepackage{authblk}
\usepackage{epigraph}
\usepackage{hyperref}
\usepackage{graphicx}
\usepackage{mathtools}   % loads »amsmath«
\usepackage{amsfonts}
\usepackage{appendix}
\usepackage{caption}
\usepackage{orcidlink}
\graphicspath{ {./figures/} }
\hypersetup{
    colorlinks=true,
    linkcolor=blue,
    urlcolor=blue,
}

\usepackage[style=apa,url=true,sorting=nyt,natbib=true]{biblatex}%<- specify style
\providecommand{\keywords}[1]
{
  \small	
  \textbf{\textit{Keywords---}} #1
}

\title{{Time is Order\footnote{Forthcoming in \textit{Time and timelessness in fundamental physics and cosmology}, eds. Silvia de Bianchi, Marco Forgione, and Laura Marongiu, Springer}}}

\author[1,2]{Álvaro Mozota Frauca}
\affil[ ]{alvaro.mozota@uab.cat, \orcidlink{0000-0002-7715-0563} \href{https://orcid.org/0000-0002-7715-0563}{https://orcid.org/
0000-0002-7715-0563}}
\affil[1]{Department of Philosophy. Universitat Aut\`onoma de
Barcelona, Building B Campus UAB 08193 Bellaterra (Barcelona), Spain}
\affil[2]{LOGOS, University of Barcelona, Department of Logic, History and
Philosophy of Science, Carrer de Montalegre 6 08001 Barcelona, Spain}

\begin{document}

\maketitle

\begin{abstract}
In this paper I argue that the fundamental aspect of our notion of time is that it defines an order relation, be it a total order relation between configurations of the world or just a partial order relation between events. This position is in contrast with a relationalist view popular in the quantum gravity literature, according to which it is just correlations between physical quantities that we observe and which capture every aspect of temporality in the world, at least according to general relativity. I argue that the view of time as defining an order relation is perfectly compatible with the way general relativity is applied, while the relationalist view has to face some challenges. This debate is important not only from the perspective of the metaphysics of space and time and of how to interpret our physical theories, but also for the development and understanding of theories of quantum gravity.
\end{abstract}


%In this paper I argue that the fundamental aspect of our notion of time is that it defines an order relation, be it a total order relation between configurations of the world or just a partial order relation between events. This position is in contrast with a relationalist view popular in the quantum gravity literature, according to which it is just correlations between physical quantities what we observe and which capture every aspect of temporality in the world, at least according to general relativity. I will argue that the view of time as defining an order relation is perfectly compatible with general relativity, while the relationalist view has to face some challenges. This debate is important not only from the perspective of the metaphysics of space and time and of how to interpret our physical theories, but also for the development and understanding of theories of quantum gravity.

\section{Introduction}

Time, or spacetime, is a fundamental ingredient of our physical theories. However, research in quantum gravity has been taken to suggest that time may not be fundamental and some authors \citep{Rovelli2011} go as far as to claim that one can dispense of time altogether. These claims are partially based on some relational views of classical physics in general and of general relativity in particular according to which all the physical content of classical theories is captured just by correlations between physical observables. In this paper I argue against this view by offering an alternative characterization of time, which I will argue captures better the notion of time of classical theories and that it is able to accommodate some physical predictions that bare correlations cannot accommodate. This view is that time is fundamentally an order relation between events.

I will start in section \ref{sect_non-relat} by reviewing the notion of time in non-relativistic physics and by arguing that time plays two roles there: it defines an absolute order relation between configurations of a system/the world and it has a metric aspect, i.e., it defines a duration. I will briefly review relationalism as is standardly understood: as a position that denies or is skeptical about the metric aspect of time. I will introduce the family of models defined by Jacobi's action as exemplifying relationalists intuitions, and I will argue that this sort of relationalism preserves the order relation.

Then, in section \ref{sect_relationalism_non-rel} I will introduce radical relationalism as a more radical version of relationalism that would even deny the physical significance of the order structure of time. Several authors in the quantum gravity community seem to hold this position, motivated by a particular analysis of the diffeomorphism invariance of general relativity and by the structures in several quantum gravity approaches. This position is applicable also to non-relativistic physics, where the symmetry of the Jacobi action mimics the symmetry of general relativity, and a similar analysis could be performed. For this and other reasons, radical relationalism also applies to non-relativistic physics. In section \ref{sect_relationalism_non-rel} I argue against radical relationalism by insisting in my positive characterization of the temporal order structure of non-relativistic models: it is indispensable for both our understanding of these models and for the physical predictions we extract from them.

Next I will move to the case of general relativity in section \ref{sect_relativistic}. I will argue that the temporal structure is much more complex and subtle in this theory, but that we can identify the same two roles of time. First, general relativity defines a proper time, which is the time an ideal clock would measure if it followed a certain trajectory in spacetime. As in the case of classical physics this metric aspect can receive some skeptical arguments or one can argue that it is relational as at the end of the day we use physical clocks made of matter for `measuring' this aspect of time. Second, in general relativity we also find an order relation, which corresponds to the causal structure of the theory. As I will argue, it is more complicated than in the case of classical mechanics as in this case the order is a partial order, but it is still an essential part of a general relativistic model. I will argue that, similarly to the case of classical mechanics, correlations between observables may fail to capture this order relation, and hence that the radical relational views of time can be challenged on these grounds.

Finally, I will briefly comment on how this position affects the foundations of quantum gravity in section \ref{sect_quantum_gravity}. First, I will argue that the relationalist positions claiming that the problem of time of some approaches to quantum gravity can be overcome can be challenged in light of what I argue in this article. And second, if spacetime functionalism is the way of making sense of spacetimeless approaches to quantum gravity, it is precisely temporal order the essential function of spacetime to be recovered by functional analysis.

\section{Time in non-relativistic physics} \label{sect_non-relat}

Most of our intuitions about time and temporality are captured by the models of classical theories. At the same time, the temporal structure of these models is less controversial, and for this reason I will start by reviewing the temporal structure of non-relativistic theories and I will argue that time plays two roles in these theories: it defines an order relation and it defines the duration of temporal intervals. I will be taking Newtonian mechanics as a paradigmatic example of non-relativistic theory, but the notion of time I will be discussing extends to many other domains in physics and science. I will also introduce relationalism as a position about time that denies the importance of the metric aspect of time, while keeping the order structure.

In Newtonian mechanics, at least in the way Newton formulated it, time and space are taken to be absolute: there is an absolute space in which bodies move and an absolute time flowing uniformly and independently of material systems. We can model Newtonian time as a real-valued parameter $t$. Each value $t_0$ identifies an instant in absolute time, a value $t_1$, $t_1>t_0$ identifies a latter instant, and the difference $t_1-t_0$ gives the duration of the interval between both instants, i.e., it measures how much time has elapsed. In this sense, we can see absolute time as playing two roles: it defines an order relation between instants and it also defines a duration for intervals. In this paper I will argue that these two roles of time are still present, even if modified, in our best physical theories, including general relativity.

A part of my claim that may be controversial is that speaking about an order relation may suggest that I am implying something like that there is a difference between past and future, that the present is in some way special, or that there is an arrow of time\footnote{Similarly, the order relation of time can be connected with other philosophical issues such as our perception of time our the notion of causality. I will also take a neutral stance with respect to those topics.}. Here I do not want to make or deny any such claim, and I want to remain neutral with respect to those debates. By order relation I just mean that in non-relativistic physics there is a fact about what happens before and what happens after. For instance, Newtonian physics predicts the positions of Mars and Venus the next 10 times the Earth is at its perihelion. This prediction is that of an ordered set, and there are facts about the positions for the first time, for the second time, and so on. In Newtonian physics, the same set but with the order reversed would also be a valid model which would define an alternative order relation. Here I want to remain neutral about whether one of the two is privileged or about whether the two are just equivalent representations of the same physics, and I want to insist that time is directly related to the ordering of events\footnote{Some authors prefer referring to the ordering structure of time as defining a `betweeness' relation rather than an order relation in order to make clear that there is no privileged temporal direction in the formalism.}. In this sense, while one can take different positions about temporal asymmetry, the fact is that the absolute time parameter in Newtonian physics naturally defines a sequence of events.

Absolute time is a notion of time that is quite strong, and relationalism appeared as a philosophical position that advocated for a less charged notion of time. Relationalists argued that one can have a picture of the world in which time is relational, i.e., not absolute, while keeping all the empirical content of Newtonian physics. Relationalist arguments criticize absolute time on the grounds that it is unobservable. For instance, Leibniz argued that a world and a copy of it in which everything happened 1 second later would be indistinguishable, and he denied that they could represent genuinely different possibilities. In this sense, Leibniz argued for a temporal relationalism according to which it does make sense to locate instants with respect to each other but not in absolute terms.

Contemporary relationalist positions go a step beyond, and they also deny the metric aspect of time\footnote{This sort of view was famously defended by Ernst Mach \citep{Mach1883}, who inspired Albert Einstein in his formulation of general relativity. More recently, and in relation to the debates surrounding quantum gravity, this view has been defended by Julian Barbour \citep{Barbour1994}.}. The basic intuition for denying the metric aspect of time is that we never measure time directly. Instead, what we use for keeping track of time are physical clocks. One could argue that when we say that one year in Venus lasts around 225 days what we are really saying is that in the time that Venus completes an orbit around the Sun, planet Earth has spun 225 times around its axis. Alternatively, we could translate those 225 days into oscillations of our most regular atomic clock, or into some other process in our favorite physical clock. In this sense, what we measure is not the absolute duration of processes, but its duration relative to other physical processes. This motivates the relationalist to deny the metric aspect of time.

Let me introduce Jacobi's model of Newtonian systems as a class of models that capture well the intuition of the relationalists and which has a strong similarity with general relativity. I will skip the technical details here and focus on the conceptually interesting part of this kind of model. I refer the interested reader to \citep{Barbour1994} and \citep{Gryb2010} for complete introductions to this kind of model. The Jacobi's action for any Newtonian system with energy $E$ is given by:
\begin{equation}
S[q]=\int d\tau \sqrt{T(q,\dot{q})(E-V(q))} \, ,
\end{equation}
where $T$ and $V$ represent the kinetic and potential energy of the system as functions of the degrees of freedom of the system $q$ and their velocities. When one finds the trajectories that extremize this action what one finds are trajectories $q(\tau)$ which are very similar to the trajectories $q(t)$ one would find making use of the standard Newtonian machinery. One can see that both $q(\tau)$ and $q(t)$ describe the same sequences of events, for instance they may describe the trajectory of a football after being kicked. However, the values of $t$ and $\tau$ at each moment are (generally) different. Moreover, Jacobi's action admits infinitely many solutions to its equations of motion which are equivalent, i.e., they describe the same sequences of configurations, but expressed as different functions $q(\tau), q(\tau '), q(\tau ''),$ etc. In this sense, while Newtonian dynamics describes how systems evolve with respect to absolute time $t$, the dynamics of Jacobi's action describes how systems evolve, but with respect to an arbitrary (monotonic) parameter $\tau$.

We can build a Jacobi model describing our solar system to illustrate the relationalist intuition in the example above. The Newtonian model tells us that Venus describes an orbit around the Sun between time $t_0$ and time $t_1=t_0+225$ days, and that in this time Earth spins 225 times around its axis. If we build a Jacobi model for describing the same system, it also predicts that there is an instant $\tau_0$ and another $\tau_1$ in between which Venus turns around the Sun and in between which Earth spins 225 times. What makes the Jacobi model different is that $\tau_0$ and $\tau_1$, or their differences, are not fixed. That is, while in every Newtonian model describing the system we necessarily have $t_1-t_0=225$ days, in the Jacobi model the difference $\tau_1-\tau_0$ could be any positive real number, as there is an infinite number of solutions of the equations of motion of this model which describe the same sequence of configurations of the system, and each of these solutions assigns different values to $\tau _0, \tau_1$ and the difference between them. That is, it could be $225$, but it could also be $1$, $\pi$, or $10^{100}$\footnote{Perhaps the reader has noticed that while in the Newtonian model $t$ carries a unit (days), for $\tau$ I am dropping it, as it is just an arbitrary parameter.}. Now, the relationalist argues that any model is equally good for describing what is happening in the solar system, no matter if it says that $t_1-t_0=225$ days or that $\tau_1-\tau_0$ is $1$, $\pi$, or $10^{100}$. As models with different temporal parametrizations or different absolute durations are empirically indistinguishable, relationalists deny that absolute duration or the values of a temporal parametrization have any empirical meaning.

Notice that it is when we consider the whole universe that the argument takes its stronger form, while if we consider just part of the universe, then it seems harder to deny that time carries some physical information in its metric aspect. For instance, imagine that for describing the dynamics of the solar system we used two Jacobi models: one describing the motion of Venus and another describing the notion of the Earth. Each of them would describe\footnote{I am at this level of discussion ignoring the gravitational interaction between both systems.} what each of the systems does, basically to rotate around the Sun and to spin around their axes, and each of them would be using an arbitrary temporal parametrization. This means that these two models together fail to capture the relations between the two motions, i.e., the fact that Venus completes an orbit in the time in which the Earth spins 225 around its axis. To capture these relations we need to use the Jacobi model including both systems or to restore Newtonian absolute time. If we add Newtonian time back to the picture, we can compare both motions and recover the facts about how they relate. In this sense, the metric aspect of time can be understood as encoding how the duration of processes of a system relates or would relate to the duration of processes of another system. The skeptical argument about the metric aspect of time can be applied to denying that there is an absolute sense of duration, but it cannot be applied to deny that the metric aspect of time has physical meaning when applied to a subsystem of the universe, as it certainly encodes relations (real or potential) with other physical systems.

Now, independently of whether we accept the relationalist argument or not, the crucial point for my argument in this article is that it does not affect the temporal order structure of our models. That is, even if we deny that the values of $\tau$ or $t$ are relevant, it is still true that all the equivalent solutions $q(\tau)$ of a Jacobi model represent the same sequence of configurations of the system. The temporal order in each of these solutions is the same, and it allows us to predict the ordered set of 225 positions that Venus will take each time the Earth completes a turn around itself. Furthermore, we take the order in this set to be a physical prediction of our models: if we obtained the same set of 225 positions in a different order (turning around the Sun in the opposite direction for instance) that would be a wrong physical prediction. In this sense, the relationalist positions based on the study of the Jacobi action deny the metric aspect of time but keep its order role.

Jacobi models are fundamentally analogous to general relativistic models. The reason for this is that the symmetry of the Jacobi action is a one-dimensional version of the diffeomorphism symmetry group of general relativity. Different solutions of the equations of motion of the Jacobi action describe the same trajectories in configuration space, i.e., the mathematical space describing all the degrees of freedom of a system, but they disagree on the way they parametrize these trajectories. That is, they disagree on the values of $\tau$ they assign to each instant of the evolution. In this sense, instead of defining a unique $q(t)$, what Jacobi action defines is an equivalence class of solutions $q(\tau)$. One can transform from one $q(\tau)$ to any other $q(\tau ')$ in the equivalence class by performing a temporal reparametrization $\tau \rightarrow \tau '$, which is just analogous to the way one can change spacetime coordinates (or perform an active diffeomorphism) $x^{\mu} \rightarrow x'^{\mu}$ in general relativity. The fact that any choice of temporal coordinate $\tau$ is as valid as any other, supports the argument of the relationalist, but it is important to notice that the order relation is independent of this choice, as any element in the equivalence class describes the same sequence of events occurring in the same order. I will use this fact to argue that the symmetry of models like Jacobi's or general relativity cannot be used as an argument for denying the importance of the temporal order structure in our models.

\section{Radical relationalism in non-relativistic physics}\label{sect_relationalism_non-rel}

Having introduced the basic notion of time in non-relativistic physics and the two functions that time plays in this kind of theory we can discuss radical relationalism as proposed or implied by part of the quantum gravity community. I will argue that the sort of relationalism that is being proposed in the literature, at least in its strongest reading, denies the order aspect of time, and reduces physics to be contained in correlations between observables. I will argue that this view of physics is problematic from the point of view of non-relativistic physics, as there are perfectly valid non-relativistic models that cannot be accounted for if we take this view of time. This will anticipate my objection in the next section, since leaving the order aspect of time out of physics will be in conflict also with the more flexible order of relativistic physics.

I will consider that authors like Carlo Rovelli, Francesca Vidotto, and Daniele Oriti \citep{Rovelli1991,Rovelli1996,Rovelli2002,Rovelli2004,Rovelli2011,Rovelli2015,Vidotto2017,Rovelli2022,Oriti2021a,Oriti2021,Marchetti2021}  defend this sort of radical relationalism. They make claims like `The quantities predicted by the theory are values of some variables when other variables have given values.' \citep[p. 21]{Rovelli2022} (by the theory here they mean any classical theory). I will take their position to be that all that a classical theory predicts is just the values of some variables when others take some other values and nothing beyond this. Crucially for my argument, this leaves out the order in which these values are taken. This reading of the relationalism of the quantum gravity community has been done also in \citep{Thebault2012,Thebault2021} and I am unaware of any discussion by the community rejecting that this is their position. Even if in some passages there is some ambiguity in their claims\footnote{For instance, in \citep{Rovelli2004} it is claimed both that the arbitrary coordinates have no physical significance (p. 79) and that there are sequences of events (p. 74).}, and the importance that the temporal order relation may or not have is rarely if ever discussed, I believe that the most consistent reading of most of their work is that it denies the physical meaning of the temporal order relation if there is no `observable' associated to it. Here I am not interested in arguing in detail the extent to which this position is endorsed, but in arguing that it is problematic at a classical level and that it has consequences in the construction and interpretation of quantum gravity. One of the features of this position that I will challenge is that clocks are necessary for making sense of physical theories.

Let me mention that the motivation for endorsing this sort of relationalism is partly an analysis of the reparametrization invariance of general relativity using some concepts of gauge theory, most relevantly the concept of observable. In my opinion, this analysis is flawed at several points as some authors have argued in the literature \citep{Maudlin2002,Pitts2014-PITCIH,Pitts2017,MozotaFrauca2023} and it makes the correlation view of physics problematic. Here I won't discuss the details of such an analysis of gauge, and I will just argue that the view emerging from that analysis doesn't accommodate some of our notions and intuitions about time and reparametrization invariant systems and that it can be challenged. In this section I will just discuss non-relativistic systems and I will leave the case of general relativity for section \ref{sect_relationalism_general_relativity}.

When it comes to discussing Newtonian systems, this view postulates that $t$, the Newtonian absolute time, is a quantity that can be measured by making use of clocks and that the basic correlations that one observes are correlations of the form $q(t)$, where $q$ is some configuration variable. By considering $t$ to be a measurable quantity, a partial observable in their terminology, the correlation view is able to recover all the physical content of Newtonian physics. That is, we are able to translate claims from the language of the correlations $q(t)$ to any other formulation of Newtonian physics in which time is framed not as an observable but as an order of events. The correlations view is problematic when the temporal structure of a model is \textbf{not} included in the set of quantities with which one builds correlations.

\begin{figure}[ht]
\centering
\includegraphics[width=0.8\textwidth]{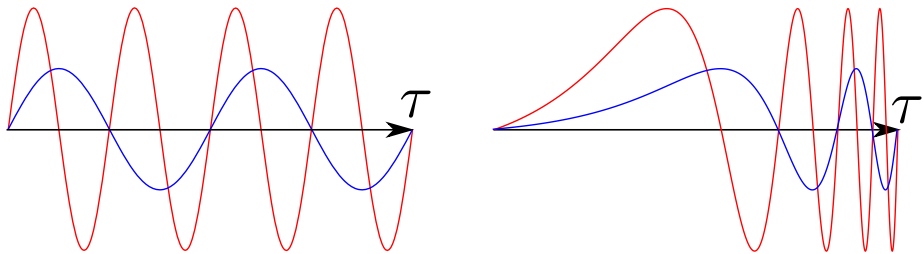}
\caption{\label{figure_oscillators} Representation of two equivalent solutions of the equations of motion of the double oscillator model. The two solutions define the same sequence of configurations but disagree on the value of the time coordinate $\tau$ they assign to each configuration.}
\end{figure}

This is well exemplified in the case of the Jacobi action of a Newtonian system as discussed above\footnote{A terminological note: Rovelli and Vidotto call this sort of system relativistic as they interpret it as a model with no preferred time variable. I prefer to stick to the traditional terminology, which in my opinion is more adequate as this model is not a model of special or general relativity, and consider this a non-relativistic model.}. Let me take the example of a system of two harmonic oscillators\footnote{This example has been studied in \citep{MozotaFrauca2023} also in connection with the problem of time of quantum gravity.}. Solutions of the equations of motion for such a model represent a trajectory in configuration space parametrized by an arbitrary parameter $\tau$. These are functions $x(\tau),y(\tau)$ as represented in figure \ref{figure_oscillators}. As discussed above, there are several ways of interpreting this model that keep as a prediction the order of configurations that any two reparametrization equivalent solutions define. In the radical relationalist view, this information is not considered part of the model and it is just the set of pairs $x,y$ along a trajectory, in no particular order, that is considered to constitute the physical content of the model. In \citep{Colosi2003,Rovelli2004} a similar model\footnote{Colosi and Rovelli fix the frequencies of both oscillators to match, which highly simplifies the trajectory in configuration space.} is built and it is argued that it is only the possible values of the position of one of the oscillator when the other takes a given value that has physical meaning, i.e., $x(y)$ or $y(x)$. Explicitly, they claim that one can build the model using a `clock' $\tau$, but that this is just a gauge parameter that one needs to `throw away'. That is, to eliminate time and keep $x(y)$ or $y(x)$ as the putative physical predictions of the model.

There is nothing inconsistent with saying that the goal of building a model is to define a set of possible observations like $x,y$. However, when we use such models we extract more information from them than just which pairs $x,y$ will obtain. As discussed above, we can also extract an order for the pairs $x,y$ from $\tau$, and even the Newtonian duration of intervals can be recovered as a preferred parametrization. In this sense, there are three possible interpretations of the model, depending on how much physical structure we read from it: we have the (not-ordered) set of pairs $x,y$, the ordered but arbitrarily parametrized trajectory $x(\tau),y(\tau)$ and the trajectory in Newtonian time $x(t),y(t)$. Rovelli and Colosi invite us to prefer the first option, and it has been further argued that gauge invariance forces one to take it, but this is just a bad application of the concepts of gauge theory beyond its domain of applicability\footnote{I refer the reader again to \citep{Maudlin2002,Pitts2014-PITCIH,Pitts2017,MozotaFrauca2023} for discussions of why the concepts of gauge theory are not readily applicable to general covariant models.}. I will argue that the most natural interpretations of this type of model, and the most common ones, are the second and third, as they incorporate facts about order that we consider physical.

Imagine that we are using the double harmonic oscillator model to model precisely a set of two harmonic oscillators that we have in a lab and that we can observe and experiment with. Then, it is natural to ask what the position of one oscillator will be the next time the other one reaches its maximum amplitude. In this context we are not interested in the whole set of values of $x$ that are taken when the other oscillator takes a position $y_0$, but just in the \textit{next} one. Similarly, for the Venus example we are interested in the succession of positions that Venus will take each time the Earth spins, as it will be compared with the succession of observations we will make from Earth. From a practical point of view, we clearly take the order that the temporal structure of the model defines to be meaningful.

The radical relationalist could argue that the order relations that a model defines are meaningful only in as much they encode correlations with an external observer or measuring system. In this way, they would translate the claim that $x(\tau_1)=x_1,y(\tau_1)=y_1$ happens before $x(\tau_2)=x_2,y(\tau_2)=y_2$ into the claim that when the double harmonic oscillator couples with an external system with configurations represented by $w$ the correlations $x_1,y_1,w_1$ and $x_2,y_2,w_2$ are allowed, with $w_2>w_1$. The relationalist can try this kind of move, but in doing so they would be acknowledging that we many times model just a part of the degrees of freedom of the universe and they would have to acknowledge that it is not just what happens to a system that we model, but also when it happens, even if they believe that this \textit{when} just encodes a relation with the rest of systems of the universe. This is in tension with the attitude held by the quantum gravity community, as they explicitly claim that they do not aim to make complete models of all the degrees of freedom of the universe, but just of part of it. This can be seen for instance in \citep{Vidotto2017}, when Vidotto claims that cosmology is different from `totology', the science that would aim to describe every single detail of the universe. If the authors considered agree that the model does not aim to cover every degree of freedom of the universe and that correlations with other systems can be encoded in the order relation, then they shouldn't dismiss the temporal order as completely unphysical.

An example where this clearly shows up is in the study of cosmological models. For many models one can leave the relativistic subtleties aside and consider models which are effectively non-relativistic and temporally reparametrization invariant. The most simple of them describes just one degree of freedom, the scale factor $a$ which roughly gives the size of the universe, evolving with respect to an arbitrary parameter $\tau$. One can read this sort of model as describing how a universe expands, contracts, or undergoes several expanding and contracting phases. There is even a way of defining a metric time, that corresponds to the proper time that some observers in the universe would measure. In this sense, this is a perfectly fine model from the perspective of the defenders of time as order, of the defenders of time as order plus duration, and also from the perspective of the cosmology community, that uses this model routinely in introductions to cosmology. Radical relationalists inspired by quantum gravity on the other hand, claim that in order for this sort of cosmological model to make sense, one needs to introduce some other degree of freedom in the model that would act as clock\footnote{This is explicit in \citep[Sect. 15.3.2]{Vidotto2017} and \citep{Marchetti2021}.}. The basic reason for arguing for this supposed need is that you need at least two variables to speak about correlations, and if there is only one variable in the model one simply cannot build them. In this sense, the correlations view is of more limited application than the view that temporal order is an important feature of our physical models, at least of the non-relativistic ones.

As I said above, the relationalist could take the basic model with one degree of freedom to encode relations with external systems in the temporal order of the model, however, they do not do so and they claim that it is necessary to explicitly add other degrees of freedom to the system in order to make sense of it. In this sense, it seems to me that they are rejecting a plausible interpretation and a respectable model and getting one step closer to the `totology' they wanted to avoid.

I have argued that from an external perspective, temporal relations can be considered physical and meaningful in a sense that does not seem in contradiction with the views of the relationalist if we interpret them as encoding relations with another system or observer. This already supposes a challenge for the relationalist. But one can also challenge the relationalist from an internal perspective. That is, we can build a model that represents a whole universe with nothing external to it or with no degree of freedom left out of the model and still claim that the temporal order of the model is meaningful. For instance, one can consider that the cosmological model describing how $a$ varies is perfectly fine for describing an empty universe with its size fluctuating or that the model of the double harmonic oscillator represents a universe containing just two particles oscillating. For starters, these models and interpretations seem perfectly fine and consistent, and the relationalist would need to provide some good reason for not considering them as physical possibilities.

The arguments by the relationalists are very tightly connected with epistemology and with the way observers perform observations. In this sense, the empty universe model is considered physical only when another system is included in the model to `measure' how the universe expands and contracts. For someone with realist inclinations, rejecting this model for the case in which no other degree of freedom may sound like rejecting that a tree falling in a forest with no one around makes a sound. Epistemological consideration can constrain our metaphysics of science, but one can suspect that the relationalist has taken it too far. Similarly, for models including human beings one can also raise skeptical arguments questioning the order relation. For instance, given my memories and records of yesterday, I can hold the belief that yesterday was real, that it happened before today, and that it had an influence on the events and states of affairs of today. But if the world was just an unordered set of things that happen, as the relationalist seems to suggest, or if they happened in an arbitrary order, I would also have access to the same state of the world as it is now, and I would also hold the same beliefs about yesterday. However, one could push this argument a bit further and raise the question of whether the instant of yesterday was or is in some sense real. The relationalist may want to deny that it happened before today, but they still want to claim that it happened or happens. In this sense, one has to be careful with skeptical arguments, as they can push one too much into denying aspects of our physical or metaphysical theories that we may not want to deny.

One can draw an analogy with properties like mass or charge. An argument can be formulated for denying that they are physical properties that we can measure, as one can argue that what we measure is the position of particles or bodies and that from their motions we infer their properties. One can reasonably believe that even if these properties are not directly measurable, they play a role in our understanding of the world and that we should consider them to be physically relevant. Similarly, even if one does not have a way of directly measuring temporal order, it can be argued that it plays a role in our theories and worldview and one can resist skeptical claims denying its importance.

If the relationalist pressed on this point and asked what role the order structure of time plays in our theories and worldview, one naturally would think in terms like evolution, causality, or locality. None of them is uncontroversial, and I do not want to enter into the details of the arguments concerning them, but it seems that our (non-relativistic) theories relate what happens at one instant with what happens immediately after and immediately before by means of the equations of motion of our theories. Of course, the equations of motion also relate temporally distant instants, but the influence of one instant can be seen as being continuously carried from one instant to another, and the form of the equations of motion suggests this. Similarly, the notion of physical influences propagating locally in space but also in time is ubiquitous in physics, and this supports the claim that the order structure of time plays a crucial role in our understanding of the world. Notice also that relationalists like Rovelli make use of the same equations of motion relying on the ordering structure for deriving their predictions in the form of correlations, even if they later on throw the order structure away. One could suspect that when they are doing so they are also throwing away important information about the world that allowed them to make the predictions in the first place.

Let me conclude this section by insisting that radical relationalism that throws away the temporal order structure of non-relativistic models may be throwing away physically relevant information about the system when doing so, and that, for the case in which the model is considered externally, I believe they would be forced to accept it, as the order structure of the model is related with facts about how it interacts with other systems. In the next section I will expand this conclusion to the case of general relativity, which has a more complex temporal structure but similar reasonings will apply.

\section{Time in general relativistic physics} \label{sect_relativistic}

\subsection{The temporal order relation in general relativity}

At first sight, the discussion above may seem irrelevant to a defender of the correlation view. After all, their view is motivated by general relativity and quantum gravity and I have been discussing non-relativistic models. However, I will argue that the arguments above can easily be extended to the case of general relativistic case. In this subsection I will argue that in general relativity both roles of time, the order and the metric roles, are present, and that as in the non-relativistic case one can raise skeptical arguments about the metric role, but that the order role seems more fundamental. I leave for the next subsection the explicit criticism of the correlation view as applied to general relativity.

In relativity theory (both special and general), the split between absolute space and absolute time of Newtonian physics is broken and instead one has spacetime. In general relativity, spacetime is dynamical in the sense that depending on the configuration of matter (and even this does not fix spacetime) one can have different spacetimes. However, I will argue that even if spacetime is a more sophisticated structure than just space and time, it still hosts the two roles of time in non-relativistic physics.

The notion of simultaneity in relativity does not play a role, and even if one can divide spacetime into space and time, there are many ways of doing so, and none seems to be physically significant in any meaningful way. In this sense, one doesn't speak about instants in relativistic physics but of events or spacetime points. Even if in relativity theory there is no physical structure defining whether two points are simultaneous, the theory incorporates an order structure that allows us to say that certain events happen before or after other events. This is the light-cone or causal structure of the theory, which distinguishes between space-like, time-like, and light-like pairs of points. When a pair of points is time-like or light-like one can say that one happens before another, and this defines a partial order relation. That is, the causal structure of the model tells us for any two points $p$ and $q$ whether $p$ is in the past of $q$, in its future, or neither\footnote{For the case of spacetimes with closed timelike curves a point $p$ can be both in the past and in the future of another $q$, I won't be explicitly dealing with this sort of spacetimes, although the local order relation is perfectly fine and unproblematic. The definition of causal structure I am using assumes that spacetimes are time-orientable, which is a widely held requirement. For a discussion of some of these subtleties I refer the reader to \citep{Manchak2011}.}. Bodies and particles in spacetime follow non-spacelike trajectories, i.e., we can see the history of every body as going from its past to its future as defined by the light-cone structure of the theory. Similarly, the laws of relativistic physics respect this causal structure: charged particles generate an electromagnetic field in the points in its future light cone, to predict the state of the world in a region one needs to know what happened in its past, and so on. 

As happened in the case of non-relativistic physics, for many physical applications this causal order may be associated with something like an arrow of time or a future-past asymmetry. In this article I want to remain neutral about this possibility, as there is nothing in the formalism of special or general relativity that indicates a preference about this issue. For any spacetime, there are two possible orientations one can choose for the partial order relation, and one could argue that only one is right and that physics should reflect this asymmetry. As in the case of non-relativistic physics, my claim that there is a partial order relation is compatible with views that accept and that deny some sort of temporal asymmetry.

The metric aspect of time is also present in relativistic physics. Given any time-like trajectory in spacetime, relativity defines a time that parametrizes the trajectory, the proper time, that is interpreted as the time an ideal clock would measure if it followed that trajectory. This time is trajectory dependent, which means that two observers following two different trajectories in spacetime that meet at two points will measure different times between both events. This gives rise to the famous twins' paradox and other relativistic effects, it has been empirically corroborated, and it is an effect that is relevant for the correct working of devices like the GPS.

As in the case of non-relativistic physics, one can formulate skeptical arguments for denying the physical relevance of this metric aspect. Proper time is defined as the time an ideal clock would measure, but we do not have ideal clocks, what we have are just physical clocks. In this sense, if our theory was able to incorporate all the degrees of freedom of the universe, including the ones that constitute our clocks, why would we need this time that we cannot measure? As in the non-relativistic case, retaining what happens in spacetime and preserving the non-metric structure of spacetime may be enough to make sense of relativistic models. We can even bring the example of the Earth and Venus to a relativistic setting: in a relativistic model of the solar system we can also make claims like that in the time Venus does an orbit around the Sun, Earth spins 225 times around its axis.

The analogy with the non-relativistic system also forces us to the conclusion that one could deny the physical relevance of the metric of time only when all the degrees of freedom of the universe are considered. When taking the perspective that some degrees of freedom are left out, proper time can be seen as encoding physically meaningful predictions for systems related to spacetime or the material systems considered by the relativistic model. For instance, even if a relativistic model describing a spacetime does not describe or take into consideration facts about a space rocket traveling in spacetime, it can be used for computing its proper time, which would be a good approximation to the readings of the clocks in the rocket.

In this sense, relativistic models are not that different from Newtonian models. Time still plays an ordering role, as one can still make claims like that the event of me having breakfast tomorrow is in my future and that the Big Bang singularity was in my past. Time still plays a metric role, as there is some sense in which one can say that from the big bang to the moment in which I am typing this 13.8 billion years have elapsed and that from this moment to tomorrow's breakfast my watch will advance a few hours. Things are slightly more complicated, as now we have to accept that simultaneity is lost, that the readings of clocks depend on their trajectories, and that the order structure is not independent of what matter does. But we still have a temporal structure playing similar roles.

I believe that from a conceptual point of view one cannot deny that both elements, order and metric, are present in relativistic models. However, it is when one discusses the diffeomorphism invariance of general relativity that confusion is prone to arise. General relativistic models are defined making use of the structures of differential geometry as a triple $\langle M,g,\phi \rangle$, i.e., a manifold $M$, a metric tensor $g$ defined on it and the matter content of the universe, represented by $\phi$. Given one triple, we can identify events with points on the manifold and the metric tensor defines the temporal relations between them. For instance, the event of me typing this can be represented by a point $p$ and the event of tomorrow's breakfast by a point $q$. My trajectory in spacetime is represented as a trajectory on $M$ and the metric tensor on the manifold determines that this trajectory is timelike, i.e., that breakfast is in my future and its duration of some hours. The diffeomorphism invariance of the theory means that one can define an equivalent triple $\langle M,g',\phi' \rangle$ in which $p$ and $q$ do not represent the same events, as it is $p'$ and $q'$ which represent me typing and my breakfast tomorrow. However, using this alternative model one can also read that my breakfast is a few hours in my future by reading the information encoded by $g'$ in the set of points that represent my trajectory in this model. In this sense, any diffeomorphism-related triple $\langle M,g,\phi \rangle$ contains the same physical information and one can claim that it is the equivalence class of triples under diffeomorphisms that really constitutes the relativistic model. This symmetry of the theory is what motivates relationalist claims like the ones I have introduced in the previous section.

Relativistic models are analogous to the Jacobi model for Newtonian systems. Recall that for the double harmonic oscillator system we claimed that all parametrizations of the same trajectory in configuration space represented the same physics. This can be formulated as saying that it is in the equivalence class of trajectories $x(\tau),y(\tau)$ that one has to extract the physics of the model. As I argued above, all the trajectories agreed on assigning the same order to the configurations that the system takes, and it is reasonable to take it to be part of the physical content of the model. For the case of relativistic physics exactly the same reasoning applies: every triple $\langle M, g, \phi \rangle$ in the equivalence class represents the same events with the same partial order relations holding between them, and one can consider them to be physical. That is, it is a prediction of a general relativistic model describing the universe surrounding me not just that I am typing and that I will have breakfast, but also that the latter is in the future of the former, and that my clock will advance a few hours in between both moments.

To conclude this subsection, let me insist on its main message: in relativistic physics the temporal structure is more sophisticated and it is even dynamical in the case of general relativity, but we find that it plays the same two roles as it still defines a (partial) order relation between events and it predicts the readings of clocks. The fact that the models are expressed using manifolds and coordinates does not change the fact that these temporal functions are well-defined. In the next subsection I turn to analyze how the radical relationalist interpretation is applied to general relativity and the way it is still affected by my arguments in section \ref{sect_relationalism_non-rel}.

\subsection{Radical relationalism in general relativity} \label{sect_relationalism_general_relativity}

As I commented in section \ref{sect_relationalism_non-rel}, the radical relationalism that is advocated for by the quantum gravity community may be too radical and throw away part of the content of our physical models that we may regard as physical. In this section I will argue that the same arguments expand to the way radical relationalism is applied to general relativity.

First, I noticed that when discussing what observables count as legitimate observables, Rovelli and Vidotto took that the Newtonian time counts as an acceptable observable in the context of Newtonian dynamics. Why shouldn't we extend this to the case of general relativity? Why shouldn't we incorporate the temporal structure of the model as an observable? The temporal structure of general relativity is more complex than the one of Newtonian mechanics, but both seem equally observable or unobservable. In the same way that Rovelli and Vidotto accept that $t$ can be considered an observable in Newtonian mechanics because one can measure it with a clock, why not accept $\tau$, the proper time along some time-like curve to be an observable for the same reason\footnote{In \citep[pp. 51-52]{Rovelli2004} it is discussed a solar-system example in which proper time is included as an observable but just when a clock is explicitly added to the system. Here I am proposing to include the proper time along any time-like curve to be considered an (partial) observable of the theory.}?

A similar comment could be made about coordinates. There are spacetimes for which one can undoubtedly construct systems of coordinates that are closely related to the temporal structure of the spacetime. The clearest example is Minkowski spacetime, where the standard coordinates can be interpreted as the proper times that some ideal clocks would measure in exchanging light pulses among them. Taking this coordinatization seems physically meaningful, and there seems that there is no reason not to take it as a partial observable, if we accepted that in a Newtonian theory $t$ could be taken as a partial observable. If one accepts one set of coordinates as meaningful and observable, one could even argue that any other set of coordinates should be considered on the same footing, as any function of a set of observables could also be considered equally observable. For more general spacetimes, engineering a way of associating spacetime coordinates to something like ideal clocks can be more complicated, but there are ways of doing so, at least for some spacetime regions. An ingenious example of this was developed precisely by Rovelli in \citep{Rovelli2002a}. This idea required four clocks moving in spacetime and sending signals into space indicating the proper time at which they emitted the signal. From any point in at least some region of spacetime one receives the four signals and this defines a coordinatization for spacetime. Rovelli baptized these coordinates as the `GPS observables' and claimed they are legitimate partial observables for the model when these four clocks and the signals they send are included in the model. But why would we limit this to just the case in which we have explicitly added the clocks and signals to the system?

This brings us back to my objection in section \ref{sect_relationalism_non-rel} that radical relationalism goes against the standard practice of physics in the sense that models are many times used in a way that encodes relations of a system with some others external to it. Even if a set of clocks filling spacetime with signals is not added to the model, one can see a model written in terms of GPS coordinates, or any other coordinate system, as encoding the same physics as the model in which we explicitly add such clocks. In this sense, even if we do not add our worldline to a solar system model or to a cosmological model, a general relativistic model predicts the succession of events that we will go through and the time that our clocks will indicate. 

Similarly, radical relationalism in general relativity has motivated claims like that we should include clocks and rods as internal degrees of freedom of our theories\footnote{See for instance the recent discussion in \citep{Marchetti2021}.}. In this sense, it rejects the idea that a prediction for a physical quantity $O$ at a spacetime point, $O(x^{\mu})$, is meaningful if it is expressed in terms of coordinates in the manifold, and it considers it meaningful only if it is expressed as a correlation with four physical quantities that would be acting like rods and clocks, i.e., individuating spacetime points. That is, a correlation $O(\phi^a)$ would be considered a physical prediction, where $\phi^a$ represents four physical degrees of freedom such as four scalar fields. In the same way that in the case of non-relativistic physics I considered this view to be problematic, I believe my analysis extends to the relativistic case. Indeed, in the usual scientific practice one rarely adds explicitly any clocks and rods to the model, and we are able to make sense of vacuum models where no matter is added. Vacuum models in general relativity present a rich phenomenology and are widely used for describing gravitational waves, cosmological settings, or the merge of black holes among others. From this perspective, the `external' view of general relativistic models seems valid, and the temporal partial order structure of the model can be seen as encoding physical information that observers not explicitly added to the model would observe. From the internal perspective one can also raise a challenge to the radical relationalist, as one can still believe that a model describing, say, the merge of two black holes, is a physically meaningful model, even if no matter is included in the model\footnote{Let me mention that the relationalist could try to argue that radical relationalism can account for such situations too by studying correlations between gravitational observables. Let me make three comments about this line of reasoning. First, this is not a strategy followed by the quantum gravity community at the time of interpreting their models. Second, as far as I know, it seems impossible to build coordinates out of the metric and related tensors. Third, for symmetric cases like Minkowski spacetime one has spacetime points with the same physical properties, and hence one cannot build coordinates based on the metric that would distinguish them.}.

Besides these arguments, in section \ref{sect_relationalism_non-rel} I argued that the temporal order of events plays an important role in grounding many of our physical intuitions like the ones related to locality, evolution, or causality. Moreover, the differential equations describing our non-relativistic physics explicitly relate the state of the world or of a system at an instant with the state of the world in the instants just before and after. This sort of argument can be extended to the case of relativistic physics, as in relativistic physics it is generally the case that locality, evolution, or causality play an important role in our reasoning and one still has local differential equations that relate the state of affairs in each spacetime point with what happens in the points surrounding it. In this sense, the order structure plays an important role in our understanding of relativistic theories.

Finally, it is clear that in relativistic models one does not just take the empirical content of a model to be contained in a non-ordered set of multiplets $\phi,g$ corresponding to the state of the matter fields $\phi$ and (possibly) of the metric field $g$ at every spacetime point, as the order that the metric tensor, together with the manifold structure, defines is taken to be part of the content of the model. This is just as in the double harmonic oscillator model, where one takes the ordered set of pairs $x,y$ to carry physical content, and a differently ordered set wouldn't be considered the same physical possibility. In the case of the relativistic model if we assigned each combination $\phi,g$ to a random point on a manifold we could equally claim that we have built a different (and physically wrong) model.

For these reasons, I believe that the claims in the quantum gravity literature that in relativistic physics it is just correlations between observables that have physical meaning are misguided and that they leave out an important part of the physical content of our relativistic models. Time, in relativistic physics, is (at least) a partial order relation between events, and any account that leaves out this feature seems to be leaving out some important physical content of our models.   

\section{Time and quantum gravity} \label{sect_quantum_gravity}

In this last section I want to make a brief comment about the implications that the analysis above has for two relevant issues in the foundations and philosophy of quantum gravity.

First, in canonical approaches it is well-known that there is a problem of time: by applying canonical quantization methods to reparametrization invariant models one gets a quantum model with a trivial evolution equation. This has been interpreted as showing that time, as the (partial or not) order relation of the original model, does not play a role in the quantum theory. Radical relationalists argue that this is not problematic because the quantization procedure still leaves room for building some quantum operators that look like correlations, which from the radical relationalist perspective is all you need to build a physical theory. For instance, for the double harmonic oscillator case there would be operators associated with $x$ and $y$, and the hope would be to build operators $\hat{X}_Y$ that would capture the correlations between both observables. If one accepts the analysis above, one would very reasonably worry that some important physical content of the classical models (both for Jacobi models and general relativity) has gone missing and that there is something problematic about the quantization carried out. In other words, they would take the problem of time to be something serious and be skeptical about the way the radical relationalist pretends to solve the problem or argues that the problem is not a problem after all. For an analysis of the problem of time along these lines, see \citep{MozotaFrauca2023}.

Second, one may nevertheless have an open mind about the possibility of a fundamental quantum theory of gravity in which there is no time at a fundamental level. In this case, it seems that something like spacetime functionalism, as discussed in the philosophy of physics literature \citep{Lam2018-LAMSIA-2,Lam2020} would be needed for recovering the appearance of a spatiotemporal world. This amounts to saying that in order for a spacetimeless theory to be compatible with our experience, it needs to have structures able to satisfy the functions that we usually ascribe to space and time. Following the analysis here, the main function of time that would need to be recovered in order for a functionalist account to be successful is precisely the order function. In this sense, it seems reasonable to demand an explanation from the spacetime functionalist or from the quantum gravity theoretician about how the order function of time can be recovered from a fundamental level of reality that would be lacking it. 

\section{Conclusions}

In this article I have argued that time plays two major roles in our physical theories: it defines an (partial or total) order relation and it has the metric role of defining the duration of processes. I have argued that while one can raise some skeptical arguments about durations, the order role is fundamental and that it seems essential for our physical theories. In this sense, I have argued that some radical relationalist views by the quantum gravity community are problematic and I have commented that this has an impact on our understanding of the philosophy and foundations of quantum gravity. I believe that for the picture of a fundamentally spacetimeless world that nevertheless looks spatiotemporal to be appealing, the quantum gravity community should address the issues raised in this article and explain how the appearance of our world as a partially ordered set of events could be compatible with their proposals.

\section*{Acknowledgements}
I want to thank the Proteus group, Carl Hoefer, and the audiences at Buenos Aires and Belgrade for their comments and discussions. 

This research is part of the Proteus project that has received funding from the European Research Council (ERC) under the  Horizon 2020 research and innovation programme (Grant agreement No. 758145) and of the project CHRONOS (PID2019-108762GB-I00) of the Spanish Ministry of Science and Innovation.
The author has no conflicts of interest to declare that are relevant to the content of this chapter.

%

%\bibliographystyle{h-elsevier}% bibliography style
%\nocite{*}
%\bibliography{covariant}

\printbibliography

\end{document}